\documentclass[reprint,amsmath,amssymb,aps,twocolumn]{revtex4-1}
\usepackage{graphicx}
\usepackage{color}
\usepackage{amsmath}
\usepackage{mathrsfs}
\usepackage{amssymb}
\usepackage{amsthm}
\usepackage{physics}
\usepackage{booktabs}
\usepackage[dvipdfm,colorlinks,urlcolor=blue,linkcolor=blue,anchorcolor=blue,citecolor=blue]{hyperref}

\usepackage{lineno}

\begin{document}
\newcommand{\bs}[1]{\boldsymbol{#1}}
%\linenumbers
%\setlength\linenumbersep{0.2cm}

\title{Thermalization of two- and three-dimensional classical lattices}

\author{Zhen Wang$^{1}$}
\thanks{These authors contributed equally to this work.}
\author{Weicheng Fu$^{2,3}$}
\thanks{These authors contributed equally to this work.}
\author{Yong Zhang$^{1,3}$}
\author{Hong Zhao$^{1,3}$}

\email{zhaoh@xmu.edu.cn}

\affiliation{$^1$Department of Physics, Xiamen University, Xiamen 361005, Fujian, China\\
$^2$Department of Physics, Tianshui Normal University, Tianshui 741001, Gansu, China\\
$^3$Lanzhou Center for Theoretical Physics, Key Laboratory of Theoretical Physics of Gansu Province, Lanzhou University, Lanzhou, Gansu 730000, China}

\date{\today}
\begin{abstract}

Whether and how a system reaches thermalization is a fundamental issue of statistical physics. While for one-dimensional lattices this issue has been intensively studied in terms of energy equipartition for more than half a century, few work has been performed in the case of two- and three-dimensional lattices, and thus the thermalization dynamics remains unclear for more realistic lattices. In this Letter we investigate analytically and numerically the time-scaling of energy relaxation in these lattices. We show that the equipartition of energy is generally reached following a universal scheme for large enough lattices, regardless of its dimensionality, its specific lattice structure, and whether the system is translation invariant or not. Our results have practical significance in exploring the effect of high-order nonlinearities, i.e., the combining effect of multi-phonon process, in solid materials.

\end{abstract}

\maketitle

{The ergodicity hypothesis is at the foundation of statistical physics. It implies that, in the thermodynamic limit, with no matter how weak nonintegrable interactions, a system will reach thermalization. The energy equipartition, i.e., the process for the energy initially distributed among partial degrees of freedom eventually spreads over all degrees of freedom, is an indicator of the ergodicity as well as the thermalization. However, the doubt of whether a lattice can really reach energy equipartition was raised after the Kolmogorov-Arnold-Moser theorem and the Fermi-Pasta-Ulam-Tsingou recurrence \cite{Fermi1955}  discovered in the 1950s. The former proves rigorously that there exists a positive-measure set of invariant tori in the phase-space of a weakly perturbed Hamiltonian system, and the latter indicates that the energy initially assigned to a normal mode of a nonlinear lattice may return to this mode after a period of evolution. Both violate the ergodicity hypothesis, stimulating extensive and intensive studies of the energy equipartition problem that lasts up to date \cite{Izrailev1966, Benettin1984,
Benettin2008,Benettin2011,
Benettin2013,Danieli2017,
Danieli2019,
PhysRevE.51.2877,PhysRevE.60.3781,
Tsaur1996,Parisi1997,Lepri1998,
Onorato2015,Pistone2018,Lvov2018,
Fu2019PRER,Fu2019,Fu2019a,
Pistone2019,Wang2020}.

For typical one-dimensional (1D) lattices, it has been inferred from extensive studies that the relaxation time to equilibrium, $T_{\rm{eq}}$, may scale in a power-law of the perturbation \cite{Izrailev1966, Benettin1984,
Benettin2008,Benettin2011,
Benettin2013,Danieli2017,
Danieli2019,
PhysRevE.51.2877,PhysRevE.60.3781,
Tsaur1996,Parisi1997,Lepri1998,
Onorato2015,Pistone2018,Lvov2018,
Fu2019PRER,Fu2019,Fu2019a,
Pistone2019,Wang2020}.
To concrete the scaling law, analytical approaches have played an essential role \cite{Tsaur1996,Parisi1997,Lepri1998,Onorato2015,Pistone2018,Lvov2018,Fu2019PRER,Fu2019,Fu2019a,Pistone2019,Wang2020}. Particularly, based on the wave turbulence approach, it is shown that even for the short one-dimensional lattices originally studied by Fermi et al., energy equipartition sets in finally, and $T_{\rm{eq}}$ has a power-law dependence on the perturbation \cite{Onorato2015,Pistone2018,Lvov2018}. For long-enough chains, the thermalization process ruled by the $n$th-order nonlinearity of perturbation potential follows a universal scaling, i.e., $T_{\rm{eq}}$ is inversely proportional to the square of the effective perturbation strength, in the weak-perturbation region \cite{Fu2019PRER,Fu2019,Fu2019a,Pistone2019,Wang2020}. Power-law relaxation scalings imply that energy equipartition is reach with no matter how weak perturbations.

However, few studies have been devoted to two-dimensional (2D) and three-dimensional (3D) lattices. The only references available, to the best of our knowledge, are summarized below. For $2$D lattices, Benettin et al. have studied a hexagonal lattice with interaction potentials of Lennard-Jones type \cite{Benettin2005,Benettin2008a} and they speculated that the energy equipartition may occur. Wang et al. have studied the energy equipartition process among out-of-plane flexural modes \cite{Wang2018,Wang2018a} of a graphene sheet. For $3$D case, Carati et al. \cite{CARATI2019121911} have studied an ionic-crystal model and inferred that ergodicity may not occur at low temperatures. Thermalization of $2$D and $3$D disordered lattices has not been studied at all. In fact, $2$D and in particularly $3$D lattices are ubiquitous and have more realistic importance, and it is known that physical properties usually depend on the dimensionality, for example, as having been revealed for the heat conductivity \cite{PhysRevLett.78.1896,PhysRevE.57.2992,PhysRevE.61.3828,LEPRI20031,doi:10.1080/00018730802538522,lepri2016thermal,PhysRevLett.125.040604}. Therefore, the study of lattices with higher spatial dimensions  is not only  of practical importance but also indispensable for a full understanding of thermalization.

In this Letter, we show that lattices with higher spatial dimensions can generally reach energy equipartition regardless of the dimensionality, the specific lattice structure, and whether the system is translation-invariant or not. In the thermodynamic and weak-perturbation limit, we first derive the kinetic equation of normal modes by adopting the generalized Gibbs ensemble (GGE) ansatz \cite{Dudnikova2003}, and then prove the connectedness of normal-mode networks. The interconnected structure of the normal-mode network enables us to explain why the time scaling of a single $n$-wave resonance, which is inversely proportional to the square of the perturbation strength, can characterize the energy equipartition time in the network. The predicted time scaling is verified by numerical simulations with several typical lattice models, specifically the $2$D hexagonal lattice, $2$D square lattice, $3$D face-centered cubic (FCC) lattice, and $3$D simple cubic (SC) lattice. For all these models, we have also studied their mass-disordered counterparts.

Without losing generality, we adopt the polynomial potential as the integrability-breaking term in the derivation of the kinetic equation. The lattices are described by the Hamiltonian
\begin{equation}\label{eq:H2}
\begin{aligned}
  H = &\sum_{\vb*{l},\alpha}\dfrac{ M(\vb*{l})}{2}\dot{u}^2_{\alpha}(\vb*{l})+\dfrac{1}{2}\sum_{\vb*{l},\vb*{l}',\alpha,\beta} \Phi_{\alpha\beta}(\vb*{l},\vb*{l}')u_{\alpha}(\vb*{l})u_{\beta}(\vb*{l}')\\
     &+\dfrac{g}{n!}\sum_{\vb*{l},\vb*{l}',\vb*{\beta}}\Psi_{\vb*{\beta}}(\vb*{l},\vb*{l}')\prod_{j=1}^{n}\left[u_{\beta_j}(\vb*{l})-u_{\beta_j}(\vb*{l}')\right],
 \end{aligned}
\end{equation}
where $M(\vb*{l})$ is the mass of the $\vb*{l}$th particle, $u_{\alpha}(\vb*{l})$ is a small displacement of the $\vb*{l}$th particle in the $\alpha$ direction, $\alpha$ and $\beta$ represent components of Cartesian coordinate axes, $\Phi_{\alpha\beta}(\vb*{l},\vb*{l}')=\eval{\frac{\partial^2 V^{(n)}}{\partial u_{\alpha}(\vb*{l})\partial u_{\beta}(\vb*{l}')}}_{0}$,
and
$\Psi_{\vb*{\beta}}(\vb*{l},\vb*{l}')=\eval{\frac{\partial^n V^{(n)}}{\partial u_{\beta_n}(\vb*{l})\dotsm \partial u_{\beta_1}(\vb*{l})}}_0$, with $V^{(n)}$ be the interparticle interaction potential of $n$th polynomials, and a shorthand notation $\vb*{\beta}=(\beta_1,\dotsc,\beta_n)$ is adopted. $g$ modulates the perturbation strength.

In Eq.~(\ref{eq:H2}) the first two terms are integrable part which is denoted as $H_0$, the other terms are the $n$th-order nonlinearity perturbation which is denoted as $H'$. Supposing that $H_0$ has a set of integrals of motion with $N_c$ members, i.e., $I_k$ with $k=1,2,3,...,N_c$, then according to the dynamical system theory these integrals of motion commute with $H_0$ in the Poisson bracket, i.e., $\{H_0,I_k\}=0$. In the presence of perturbation $H'$, the system evolves following the Liouville equation \cite{Zwanzig2001},
\begin{equation}\label{eq:ILE}
\dfrac{\partial}{\partial t}I_k=L_0I_k+L'I_k,
\end{equation}
where $L_0$ and $L'$ are Liouville operators, i.e., $L_0I_k=\{H_0,I_k\}$ and $L'I_k=\{H',I_k\}$.

The kinetic equation up to the second-order perturbation for $I_k$ (the zero-order term vanishes always) is
\begin{align}\label{eq:kinetic_equation-1}
\expval{\dot{I}_k(t)} \approx \expval{L'I_k}_{f(0)}+\int_{0}^{t}\expval{L'L'(t-s)I_k(t)}_{f(t)}\dd{s},
\end{align}
where $L'(\tau)=e^{-L_{0}\tau} L'e^{L_0\tau}$, $f(t)$ is the distribution function of
the system (\ref{eq:H2}).  With the weak enough perturbation, the dynamics of the system is still dominated by $ H_0$, i.e., the deformed  $I_k(t)$ can still be considered as the integral of motion of $H_0$. In this case, the distribution can be approximated by a GGE \cite{Dudnikova2003}, i.e.,
\begin{equation}\label{ff}
  f_{\mathrm{GGE}}(t) = C\exp\left(-\sum_{k}\lambda_{k}(t)I_k(t)\right),
\end{equation}
approximately, where $C$ is a normalization factor. The role of the perturbation is to drive $I_k$ evolving in the space of integrals of motion of $H_0$.

For a lattice model, introducing the normal modes $P_s(k)$ and $Q_s(k)$ and defining its complex amplitude as
$a_s(k)=(P_s(k)+i\omega_s(k) Q_s(k))$, the integrable part of Hamiltonian can be rewritten as
\begin{equation}\label{eq:H0}
 H_0= \sum_{s}\sum_{k}\omega_{s}^{k}a_{s}(k)a_{s}^{*}(k),
\end{equation}
where the frequency $\omega_{s}^{k}$ obeys the dispersion relation, $s$ enumerates the set of cardinality $\{s\}$ of frequencies associated with wave vector $k$, corresponding to the number of branches of the dispersion relation. The integral of motion of $H_0$ turns to be $I_{k}^{s} = a_{s}(k)a_{s}^*(k)$ with such a notation. Similarly, the $n$th order perturbation terms can be rewritten as
\begin{equation}\label{eq:H0}
 H'=\dfrac{g}{n!}\sum_{\ell=0}^{n}\tbinom{n}{\ell}\sum_{\substack{s_1\ldots s_n\\k_1\ldots k_n}}W_{1\ldots \ell}^{\ell+1\dots n}a_{1}\dotsm a_{\ell}a_{\ell+1}^{*}\dotsm a_{n}^{*},
\end{equation}
where $a_{j}=a_{s_j}(k_j)$ is a shorthand notation, $W_{1\dotso\ell}^{\ell+1\dotso n}$ weights the transfer of energy among waves $a_1,\dotsc, a_n$.

For the perturbation (\ref{eq:H0}), we solve Eq.~(\ref{eq:kinetic_equation-1}) and it follows immediately that, for a specific integral of motion, $I_{k_1}^{s_1}$ for example,
\begin{equation}\label{eq:kenetic_equation}
  \langle\dot{I}_{k_1}^{s_1}(t)\rangle = \eta_{k_1}^{s_1} - \gamma_{k_1}^{s_1} \langle I_{k_1}^{s_1}(t)\rangle,
\end{equation}
where $\eta_{k_1}^{s_1}$ and $\gamma_{k_1}^{s_1}$ are $I_{k_1}^{s_1}$-independent constants proportional to $g^2$. As long as $\gamma_{k_1}^{s_1}$ is non-vanishing, we obtain that $\langle I_{k_1}^{s_1}(t)\rangle$ relaxes with the time scaling $\simeq 1/\gamma_{k_1}^{s_1}$. (Please see the supplementary material (SM) for the detailed derivation of the kinetic equation and the discussion about the GGE).

Non-vanishing $\gamma_{k_1}^{s_1}$ is guaranteed by the $n$-wave resonance condition (see SM for details),
\begin{equation}\label{eq:MWRC_w}
\omega_{k_1}^{s_1}+\ldots+\omega_{k_\ell}^{s_\ell}=\omega_{k_{\ell+1}}^{s_{\ell+1}}+\ldots+\omega_{k_{n}}^{s_n},
\end{equation}
and
\begin{equation}\label{eq:MWRC_k}
k_1+\ldots+k_\ell=k_{\ell+1}+\ldots+k_{n}.
\end{equation}
Note that due to the existence of nonlinearity, a frequency would be broadened by the nonlinearity \cite{2011LNP825N}, leading to a finite width proportional to  $\gamma_{k_1}^{s_1}$. As a result, the resonance condition Eq.~(\ref{eq:MWRC_w}) can be replaced by the quasi-resonance condition for a finite-size system,
\begin{equation}\label{eq:quasi-MWRC_w}
\left|\omega_{k_1}^{s_1}+\ldots+\omega_{k_\ell}^{s_\ell}-\omega_{k_{\ell+1}}^{s_{\ell+1}}+
\ldots+\omega_{k_{n}}^{s_n}\right|\lesssim {\gamma_{k_1}^{s_1}}.
\end{equation}
The frequency spacing decreases as $\sim 1/N$ for low frequency modes, or $\sim 1/N^2$ for high-frequency modes. Consequently, this condition must be satisfied for a fixed nonlinearity $g$ as long as $N$ is large enough, and $\omega_k^s$ can take continues values in a bounded interval for a lattice model with fixed lattice space.

\begin{figure}[t]
  \centering
  \includegraphics[width=1\columnwidth]{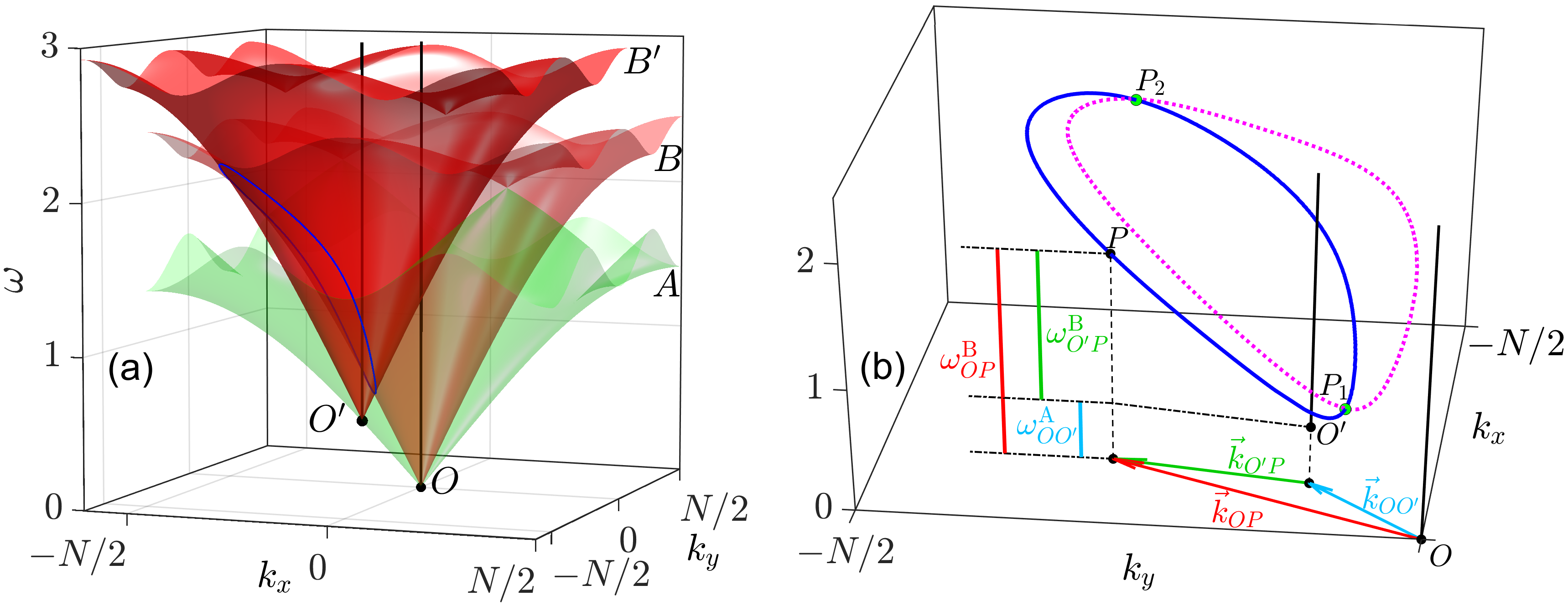}
  \caption{Graphical solution for the three-wave resonances of the hexagonal lattice. (a) The dispersion relation has two frequency surfaces denoted by $A$ and $B$. The surface $B'$ is obtained by shift the origin $O$ to $O'$ on the surface $A$. (b) The intersection curves between $B$ and $B'$ (solid line), and between the $B$ and $B''$ (dashed line), respectively. The surface $B''$ (not shown in the figure) is obtained by further shift the origin from $O'$ to $O''$ on the surface $A$. These two closed curves represent two groups of the resonant three-wave sets, which are interconnected by the intersection points $P_1$ and $P_2$.}
  \label{fig:Figure-1}
\end{figure}
For typical $2$D and $3$D lattices, the resonance conditions can be easily satisfied, as explained in the following by taking the hexagonal lattice for $n=3$ as an example. Figure \ref{fig:Figure-1}(a) shows its dispersion relation. There are two frequency surfaces denoted as $A$ and $B$. In the thermodynamic limit normal modes are dense in the surfaces, and their frequencies are continuous in the presence of nonlinearity. Given this, we search the three-wave resonant solutions following the standard algorithm \cite{1992kstbookZ}. Concretely, we move the coordinate origin $O$ to $O'$ along the surface $A$, and mark the frequency surfaces obtained after the movement as $A'$ and $B'$, respectively. Due to the topology of the surfaces, $B'$ and $B$ must have a closed curve of intersection (and so do $A'$ and $B$; the argument below can be applied to this case straightforwardly, see SM for details), see Fig. \ref{fig:Figure-1}(a). To show it more clearly, we replot the closed curve in Fig. \ref{fig:Figure-1}(b) (blue solid line). Take any point $P$ on the curve, we have $\bf{\bs k}_{OO'}+\bf{\bs k}_{O'P} = \bf{\bs k}_{OP}$ and $\omega_ {OO'}+\omega_{O'P} = \omega_{OP}$. By shifting $O$ over the surface $B$ one can show that each normal mode satisfy the three-wave resonant conditions.

Meeting the resonance conditions does not mean that energy can be transferred over all the normal modes from a specific one. We need to show that normal modes are connected as a resonant network. Note that the mode $\bf{\bs k}_{OO'}$ appears as a node of the three-wave resonant solutions, which disperses energy simultaneously to an infinite sets of modes $\bf{\bs k}_{O'P}$ and $\bf{\bs k}_{OP}$ determined by the closed curve. Then, by further shifting the origin slightly to another point $O''$ near $O'$, we would obtain another closed intersection curve (magenta dotted line) which leads to the three-wave resonant solutions involving the mode $\bf {\bs k}_{OO''}$ as the common node. Limited by the geometry of the dispersion relation, this curve must intersect with the first one at two points denoted by $P_1$ and $P_2$. Then, normal modes of three-wave resonances identified with the two curves are connected by these two intersection points. In this way, one can show that all the normal modes on the surface $B$ are connected as a network of three-wave resonances. Similar analysis can be applied to other $2$D and $3$D lattices, as long as they have multiple phonon branches.

Based on above theoretical analysis, we can outline the process of energy diffusion on the $n$th order resonance net. At the initial stage, a normal mode evolves following  Eq.(\ref{eq:kenetic_equation}) and its energy relaxes with time scaling $\simeq 1/\gamma_{k_1}^{s_1}$. This process transfers the energy of the normal mode to an infinite number of normal modes (as characterized by the closed curve in Fig.1). Then energy that flowed out would further disperse to other normal modes by a cascade of $n$-wave resonances until all the normal modes are involved, implying that a complete equipartition requires a much longer time. Since shared by an infinite number of normal modes, the energy carried by a specific normal mode after the initial stage would have become sufficiently small. Therefore, the initial stage determines the time scaling of energy dispersion, so that the relaxation time scale defined by the kinetic equation captures the time scaling of energy equipartition as well as that of thermalization. As a result, we obtain $T_{\rm{eq}} \simeq 1/\gamma_{k_1}^{s_1} $.

Note that the Hamiltonian (\ref{eq:H2}) can be rescaled by the energy density $\varepsilon$, i.e., let $u_{\alpha}(\vb*{l})={\tilde u}_{\alpha}(\mathbf{\bs l})\varepsilon^{1/2}$ we have ${\tilde H}=H/\varepsilon=H_0(\dot{{\tilde u}}_{\alpha},{\tilde u}_{\alpha})+{\tilde g} V^n(\dot{{\tilde u}}_{\alpha},{\tilde u}_{\alpha})$ with ${\tilde g}=g\varepsilon^{(n-2)/2}$. The parameter ${\tilde g}$ is the effective perturbation strength. Hence, $T_{\rm{eq}}$  can be rewritten as
\begin{equation}\label{eq:teq}
  T_{\rm{eq}} \propto {\tilde g}^{-2} = {g}^{-2} \varepsilon^{(2-n)}.
\end{equation}

The resonance condition Eq.~(\ref{eq:MWRC_k}) is for translation-invariant systems. For disordered systems, only the condition given by  Eq.~(\ref{eq:MWRC_w}) or Eq.~(\ref{eq:quasi-MWRC_w}) should be satisfied, which makes above discussion more simple.

{\bf Numerical verification}. In principle, our theoretical analysis is applicable to infinite large lattices. However, due to the quasi-resonance condition, it allows us to numerically check our theoretical predictions with finite lattices. We take the polynomial perturbation potential $ V^{(n)}=\dfrac{g}{n}\sum_{\boldsymbol{l},\boldsymbol{l'},\alpha}\left|
u_{\alpha}(\vb*{l})-u_{\alpha}\right|^n$ to check the prediction of Eq.~(\ref{eq:teq}). Faithfully, one should adopt $ V^{(n)}=\dfrac{g}{n}\sum_{\boldsymbol{l},\boldsymbol{l'},\alpha}(
u_{\alpha}(\vb*{l})-u_{\alpha})^n$ for this purpose. But for $2$D and $3$D lattices, such a polynomial potential with odd $n$ has a hyperbolic surface that makes the simulation diverge. Therefore we employ the former to perform the numerical verification. The perturbation potential with $n=3$ and $n=4$ are studied here. In our simulations, we set mass $m=1$ for particles in the homogeneous models, and random mass distributed uniformly in $(0.8,1.2)$ for particles in the disordered models.

\begin{figure}
  \centering
\includegraphics[width=1\columnwidth]{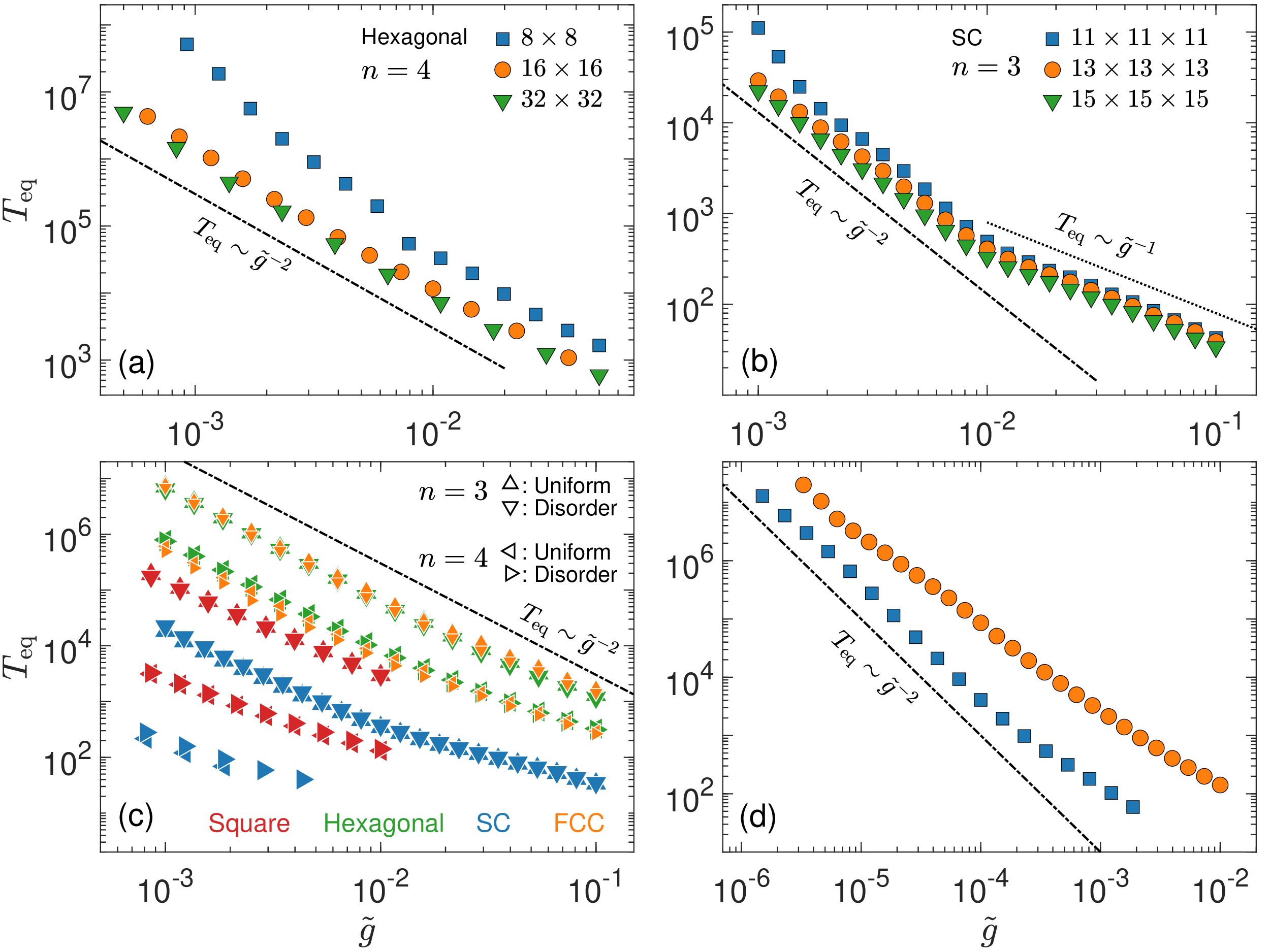}
\caption{The thermalization time $T_{\rm eq}$ as a function of perturbation strength $\tilde{g}$.  (a) and (b) show the finite-size effect for the hexagonal and SC lattices. (c) shows the results for the four lattices with larger system sizes where the finite-size effect has been suppressed. The plot (d) extends the verification to further small $\tilde{g}$ for the square and SC lattices with perturbation potential of $n=4$.}
  \label{fig:Figure-2}
\end{figure}

Figure \ref{fig:Figure-2}(a) and \ref{fig:Figure-2}(b) show the finite-size effect of $T_{\rm{eq}}$ as a function of $\tilde{g}$ for the hexagonal lattice with $n=4$ and the SC lattice with $n=3$ as examples. We see that with $32\times 32$ for the hexagonal lattice and $13\times 13\times 13$ for the simple cubic lattice $T_{\rm{eq}}$ already begins to be free of the finite-size effect within the energy region we studied.

With larger system sizes where the finite-size effect is suppressed, we show $T_{\rm{eq}}$ as a function of $\tilde{g} $ in Fig.~\ref{fig:Figure-2}(c). It can be seen that the relaxation behavior for lattices with uniform mass (up/left triangles for $n=3$, and $n=4$, respectively) and disordered mass (down/right triangles) shows neither qualitative nor quantitatively noticeable difference. For the hexagonal lattice and the FCC lattice, the prediction of Eq.~(\ref{eq:teq}) holds for both $n=3$ (up/down triangles) and $n=4$ (left/right triangles).
In particular, for the case of $n=3$, it can be seen clearly from Fig. 2(c) that the simulation results for the square and SC lattices tend to the theoretical result of the three-wave resonance as $\tilde g$ decreases, but for $n=4$, the simulation results of these two lattices seem to deviate from the three-wave resonance prediction. However, with further decreasing the perturbation strength, we find that their simulation results converge instead to the theoretical results of the four-wave resonance [see Fig. 2(d)]. Therefore, the prediction Eq.~(\ref{eq:teq}) is well verified in the weak-perturbation limit, regardless of the dimensionality, the structure of lattices, and whether the masses of particles are ordered or not.

For a real system the interaction potential usually involves various polynomial terms. Following our theoretical analysis above, connected networks of resonances of different order coexist, and in principle the energy equipartition is fulfilled through the paths in all these networks. A normal mode can disperse its energy simultaneously by three-wave resonances as well as high-order resonances. Since lattice particles vibrate around their equilibrium positions, it can be expected that the resonance dynamics is normally determined by the cubic nonlinearity in the low-temperature limit. In the finite-temperature region, however, high-order resonances may also play a role, depending on the effective strength of polynomial terms.
From Fig.~\ref{fig:Figure-2} we see that $T_{\rm{eq}}$  of $n=4$ is smaller than $T_{\rm{eq}}$  of $n=3$ at a fixed $\tilde {g}$, and the difference is much larger in the square and SC models than in other two models. This property implies that the effect of high-order resonances depends also on the lattice structure.

Figure \ref{fig:Figure-3} shows the results for the four lattice structures with the Lennard-Jones potential $V(u)=\frac{1}{72}\left[1/(1+u)^6-1\right]^2$. This potential involves all of the polynomial potential terms. In Fig.~\ref{fig:Figure-3}(a) we show $T_{\rm{eq}}$ as a function of $\varepsilon$ for the hexagonal, Square, and FCC lattices. The sizes adopted are large enough to avoid the finite-size effect. Here only the results for homogenous models are shown, but it has been verified that introducing the random masses does not change the scaling behavior, the same as seen in Fig. 2. We can see that $T_{\rm{eq}}$ follows the three-wave resonance prediction ($ T_{\rm{eq}} \propto \varepsilon^{-1} $) for all three lattices, while for the square lattice the convergence is comparatively slowly. For the SC lattice, the system size required to suppress the finite-size effect in the low-energy region has gone beyond our computing resource (With the maximum size of $17\times 17 \times 17$ we can manage to simulate, the memory capacity for only the frequency matrix has exceeded $40$GB, so that it is impractical to simulate any larger lattices for most computers). Figure~\ref{fig:Figure-3}(b) shows the results for three sizes we can access. It can be seen that in low-energy region the time scaling is close to the four-wave resonance prediction ($ T_{\rm{eq}} \propto \varepsilon^{-2} $), but tends closer to the three-wave resonance prediction with the increase of the lattice size.

The scaling behavior of the square and SC lattices indicates the effect of high-order resonances. The scaling for the square lattice is about $T_{\rm{eq}} \propto \varepsilon^{-1.3} $) in a wide region, $\varepsilon \in (10^{-5}, 10^{-4})$. For the SC lattice, though we can not make a conclusion whether the scaling may turn to the three-wave resonance prediction in the low-energy limit, it is obvious that in the region of $\varepsilon \in (4 \times 10^{-5}, 10^{-4})$, $ T_{\rm{eq}}$ has become size-independent and converges to $ T_{\rm{eq}} \propto \varepsilon^{-1.5} $. The deviation from the three-wave prediction of $ T_{\rm{eq}} \propto \varepsilon^{-1} $ should be the effect of four-wave resonances or resonances of higher orders.

\begin{figure}
  \centering
  \includegraphics[width=1\columnwidth]{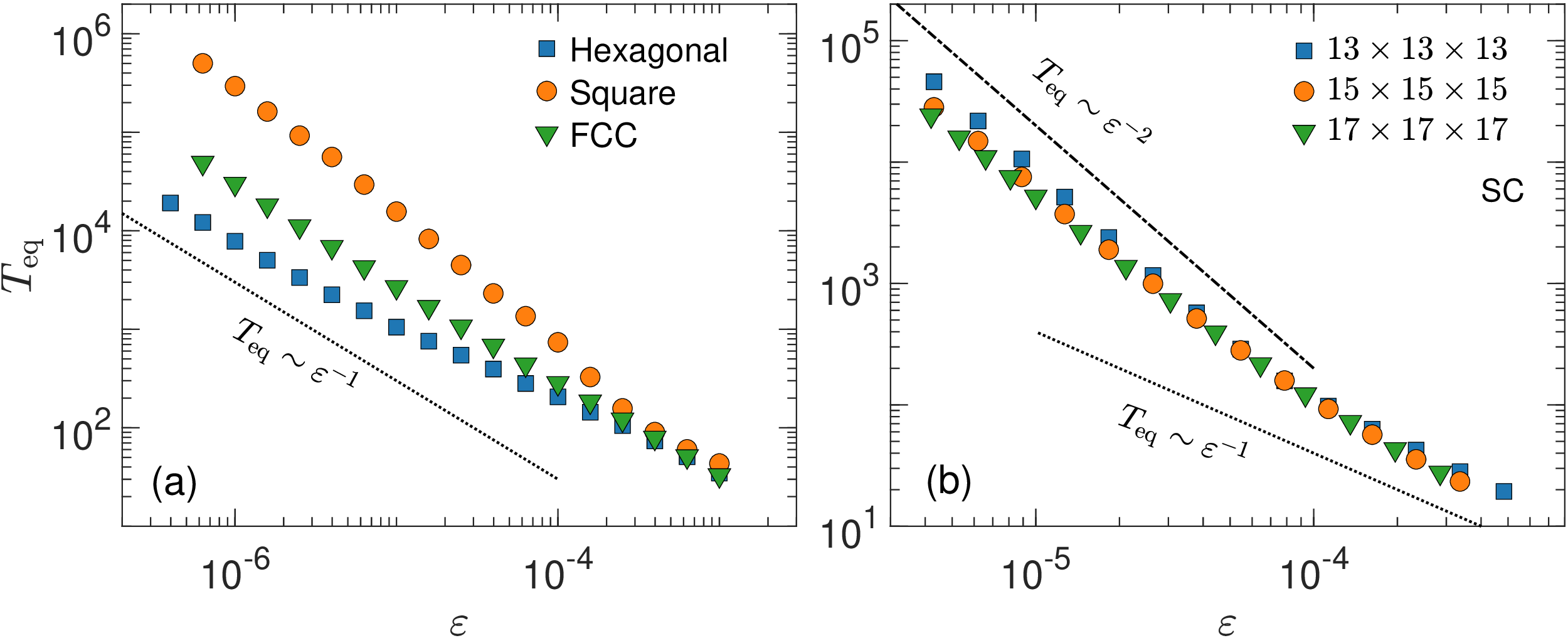}
  \caption{ $T_{\rm eq}$ via $\varepsilon$ for lattice models with Lennard-Jones potential. (a) for the hexagonal, square and FCC lattice with larger system sizes where the finite-size effect has been suppressed. (b) shows the finite-size effect for the SC lattice.}
  \label{fig:Figure-3}
\end{figure}

In summary, we have shown that typical $2$D and $3$D classical lattice models with sufficient large sizes can be thermalized in general. In a realistic lattice, networks of different $n$-wave resonances coexist. The time scaling of energy relaxation via a specific $n$-wave resonant network obeys the universal law of Eq.~(\ref{eq:teq}). The three-wave resonances may usually play the dominant roles, but high-order resonances may also play a role depending on the temperature, the interaction potential, and the lattice structure. The effect of high-order phonon processes in real materials has become a hot topic in condensed matter studies in recent years \cite{PhysRevLett.111.025901,PhysRevX.10.021063,PhysRevB.103.184302}. The time scaling of $T_{\rm{eq}}$ here can be used to identify the multi-phonon effects: $ T_{\rm{eq}} \propto \varepsilon^{-1} $ means the three-phonon dominance while $ T_{\rm{eq}} \propto \varepsilon^{-\delta} $ with $\delta > 1 $ implies a combining effect of three-phonon and multi-phonon effect.

\begin{acknowledgments}
This work was supported by the NSFC (Grants No. 11975189, No. 11975190, No. 12005156, No. 12047501); the Natural Science Foundation of Gansu Province (Grants No. 20JR5RA494, and No. 21JR1RE289); the Innovation Fund for Colleges and Universities from Department of Education of Gansu Province (Grant No. 2020B-169); the Project of Fu-Xi Scientific Research Innovation Team, Tianshui Normal University (Grant No. FXD2020-02); and the Education Project of Open Competition for the Best Candidates from Department of Education of Gansu Province, China (Grant No. 2021jyjbgs-06).
\end{acknowledgments}

\bibliography{Themalization3D_0731}
\newpage
\clearpage
\end{document}